# Automatic Generation of High-Coverage Tests for RTL Designs using Software Techniques and Tools

Yu Zhang[1,2], Wenlong Feng[1] and Mengxing Huang[1]
[1] College of Information Science & Technology, Hainan University.
Haikou, Hainan, China
[2] yuzhang.nwpu@gmail.com

*Abstract*—Register Transfer Level (RTL) design validation is a crucial stage in the hardware design process. We present a new approach to enhancing RTL design validation using available software techniques and tools. Our approach converts the source code of a RTL design into a C++ software program. Then a powerful symbolic execution engine is employed to execute the converted C++ program symbolically to generate test cases. To better generate efficient test cases, we limit the number of cycles to guide symbolic execution. Moreover, we add bit-level symbolic variable support into the symbolic execution engine. Generated test cases are further evaluated by simulating the RTL design to get accurate coverage. We have evaluated the approach on a floating point unit (FPU) design. The preliminary results show that our approach can deliver high-quality tests to achieve high coverage.

## I. INTRODUCTION

RTL design validation is a crucial stage in the hardware design process. Driven by increasing design complexity and decreasing time-to-market, RTL design validation is a significant component of the validation cost. To reduce the development cost, it is very important to develop efficient validation approaches for RTL validation.

Currently, RTL design validation mainly uses random stimuli and directed testing. Random stimuli can quickly generate many input vectors and is easy to apply to RTL simulation while facing major challenges in achieving high coverage and avoiding high redundancy in input vectors. Therefore, random generated vectors are often combined with directed testing approach. Directed testing is usually efficient in generating useful input vectors. However, developing such vectors is a labor-intensive and time consuming process. Furthermore, even a large number of vectors cannot give the engineers enough confidence since some corner cases are easily missed. Despite there have been many approaches for RTL design validation in the past decade, generation of vector sequences remains to be one of the hardest tasks in design validation and functional test generation.

Symbolic execution is a technique that has received much attention in recent years [1-3] in the context of software testing due to its ability to automatically explore multiple program paths and reason about the program's behavior along each of them. In the past several years, symbolic execution has been applied to hardware domain in different aspects: RTL analysis [4-6], high-level synthesis [7, 8], GPU program analysis [9, 10] and post-silicon functional validation [11-13].

Recently, there have been several research about developing a symbolic execution framework for RTL design [4-6]. However, it is not easy to implement a sufficient symbolic execution engine to support all RTL features. Inspired by advances of symbolic execution for software testing, we suspect if we can convert RTL designs as a C/C++ software program. If so, we can generate test cases for C/C++ software programs using the available symbolic execution tools. In this way, the tests generated for C/C++ program can be used for RTL design validation.

In this paper, we present a generic symbolic execution framework for RTL design validation (SE4RDV). SE4RDV can convert the RTL design into a C++ program. Advanced software approaches (e.g. symbolic execution) can be applied to the C++ program. In this paper, we have applied symbolic execution to generating efficient input vectors for RTL designs. To better generate efficient test cases, we limit the number of cycles to guide symbolic execution. Moreover, we add bit-level symbolic variable support into the symbolic execution engine. Generated test cases are further evaluated by simulating the RTL design to get accurate coverage.

Our research makes the following four key contributions:
 1) **Convert a RTL design validation problem into a C++ program validation problem.** A generic framework of RTL design validation is proposed. The framework converts a RTL design into a C++ program. Then a symbolic execution engine is employed to explore paths of the converted C++ program and generate tests.

The generated tests are further applied for RTL simulation and evaluate the coverage.

2) **Generate test harness.** The generated C++ program cannot be consumed by the symbolic execution engine directly. We further generate a test harness for each generated C++ program. The test harness works the same way as the test bench for RTL simulation. Moreover, we need to declare symbolic inputs in the test harness.

3) **Support bit-level variables.** For the symbolic execution engine we adapted, the minimum variable size supported is 8 bits. However, it is very common to use variables in arbitrary bits for a RTL design. We have added support in the symbolic execution engine to support defining a variable in arbitrary bits.

4) **Evaluate on practical RTL designs.** We have evaluated our approach on an OpenCores project: a FPU RTL design. The results show that our approach can achieve high coverage.

The remainder of this paper is structured as follows. Section 2 reviews related work and provides the background. Section 3 presents our SE4RDV framework. Section 4 discusses the results. Section 5 concludes and discusses future work.

## II. BACKGROUND

### A. Previous Work

Recently, there have been several frameworks proposed for executing RTL designs symbolically. In [4], STAR analyzes the source code of RTL designs statically. STAR provides a hybrid approach by mixing symbolic simulation and concrete simulation of RTL designs. In [5], HYBRO utilized software analysis techniques as a guiding metric for test pattern generation. In particular, HYBRO uses instrumented HDL, unrolls the circuit execution and uses a Satisfiability Modulo Theory (SMT) solver in order to find satisfiable assignments for conditions not seen during a previously applied execution. Due to the computational costs of SMT solvers, HYBRO is limited in the number of cycles that it can unroll the circuit. This means that branches requiring longer sequences of vectors may be completely untouched by such an approach.

In [6], a scalable approach has been proposed to enable directed test generation for RTL designs by interleaving concrete and symbolic execution. The RTL designs are instrumented first. During the simulation, the instrumented code produces a trace file. Then the trace is analyzed using a constraint solver for generating more test cases.

### B. Symbolic Execution on Hardware Domain

Recently, symbolic execution and concolic execution [14-16] technique and tools have been widely explored. Tools such as KLEE [17], SAGE [18], JPF-SE [19], S2E [20], jCUTE [21], BitBlaze [22], Pex [23] and Pathgrind [24] are just some of the symbolic execution engines currently used successfully in academia and in industry. These tools have demonstrated great advantages in software testing and security verification [25, 26].

Suppose we have a C function shown in Figure 1, the function takes two integers as inputs and return one integer. Inside the function, the input variable values are checked to decide the return value. If we want to cover all cases using concrete inputs, we need to first understand the function and then create at least four concrete test cases.

```
int function(int a, int b)
{
   int x = 0;

   if (a > 2) {
      x = 1;
   }

   if (b < 5) {
      x += 2;
   }

   return x;
}
```

Fig. 1 A simple C program.

If we run the C program using symbolic execution, we only need to execute the function once. The symbolic execution engine explores all possible paths and generate test cases for each path explored. For the function shown in Figure 1, there are four paths shown as Figure 2. For different paths, return values are different. The generated test cases by symbolic execution engine can cover all branches and paths in this function.

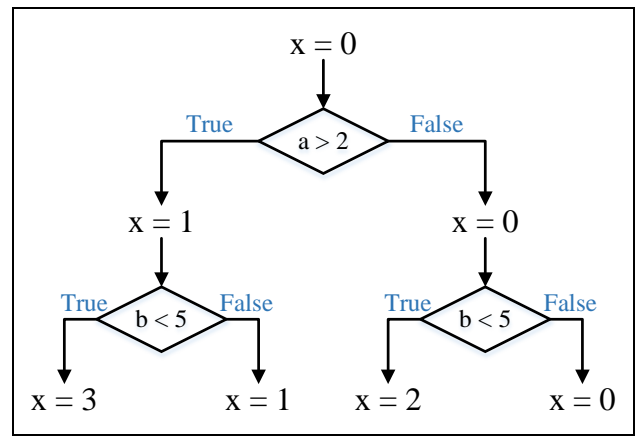

Fig. 2 A symbolic execution example.

Inspired by great success of symbolic execution on software validation, symbolic execution technique and tools have been further applied to hardware domain. For high-level synthesis, symbolic execution is used for executing both high-level C design and low-level RTL design symbolically and then equivalence checking is conducted. For post-silicon functional validation, symbolic execution of virtual devices has been employed to support test generation [11, 27, 28] and conformance checking [12, 29, 30].

In our approach, we have developed a generic framework to symbolic execution of RTL designs using software validation technique and tools.

### III. SE4RDV FRAMEWORK

#### A. Overview

As shown in Figure 1, there are mainly three steps in SE4RDV flow:
1) **Conversion**: The first step of SE4RDV flow is to convert a RTL design into an equivalent C++ program.
2) **Symbolic Execution**: The generated C++ program is executed by a powerful symbolic execution engine to explore all paths of the C++ program to generate test cases. The generated C++ program is not complete. In order to run it symbolically, a test harness needs to be created to guide symbolic execution.
3) **Simulation**: The generated test cases are applied to RTL simulation to validate the RTL design and evaluate the coverage. For RTL simulation, a test bench needs to be created. The test bench takes the generated test cases as inputs and guides the RTL simulation.

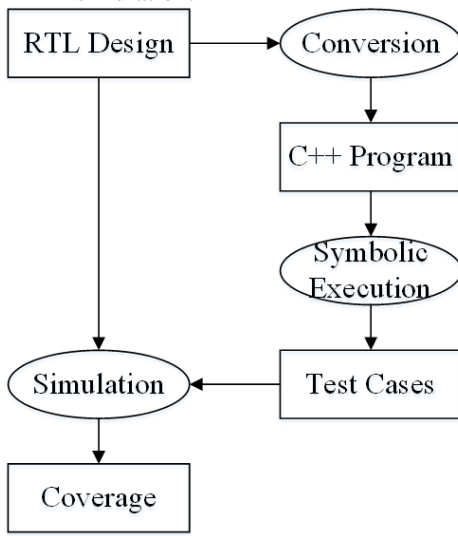

Fig. 3 SE4RDV workflow.

#### B. Illustrative Example

To better illustrate our approach, we use a simple example to show the SE4RDV workflow. The design is a simple mux implementation shown in Figure 2. In this design, the output *mux_out* is either from *din_0* or *din_1*. The branch is decided by *sel*.

```
1 module mux (
2 din_0     , // Mux first input
3 din_1     , // Mux Second input
4 sel       , // Select input
5 mux_out     // Mux output
6 );
7 //-----------Input Ports---------------
8 input din_0, din_1, sel;
9 //-----------Output Ports---------------
10 output mux_out;
11 //------------Internal Variables--------
12 reg mux_out;
13
14 always @ (sel or din_0 or din_1)
15 begin
16   if (sel == 1'b0) begin
17     mux_out = din_0;
18   end else begin
19     mux_out = din_1;
20   end
21 end
22
23 endmodule
```

Fig. 4 A simple RTL MUX design

#### C. RTL to C++ Conversion

In our approach, we convert RTL designs into C++ programs using Verilator [31]. In the converted C++ program, we basically need to focus on two parts: the inputs and outputs of the design shown in Figure 3 and the evaluation function shown in Figure 4.

#### D. Test Harness Generation

After converting RTL designs into C++ programs, we further create a test harness to guide the symbolic execution. An example is shown in Figure 5.

Since there is no clock signal in the MUX implementation, the test harness is quite simple. We only want to make the variable *sel* symbolic and run the design once to generate test cases. For the other complex designs, the test harness can be more complex. However, the idea is similar to the one shown in Figure 5.

```
VL_MODULE(design) {
 public:
   // PORTS
   VL_IN8(din_0, 0, 0);
   VL_IN8(din_1, 0, 0);
   VL_IN8(sel, 0, 0);
   VL_IN8(mux_out, 0, 0);
   ……
}

// Constructor
VL_CTOR_IMP(design) {
 ……
}

// "eval" function which is invoked each cycle
void design::eval() {
 ……
}
```

Fig. 4 The generated C++ file.

```
int main(int argc, char **argv, char **env)
{
    design* top = new design;
    bool din_0, din_1, sel;

    make_symbolic(&sel, sizeof(sel), 1, "sel");
    top->sel = sel;
    top->eval();

    delete top;
    return 0;
}
```

Fig. 5 A sample test harness.

### E. Symbolic Execution and Bit-level Support

The generated C++ programs and developed test harness are then executed by the symbolic execution engine. For the given example, there are two test cases generated.

For the available symbolic execution engine, the minimum number bits supported is 8. In the test harness, the size of variable *sel* is 1 byte. The actual generated test cases are: *sel* = 0x0 and *sel* = 0xff. But those results do not conform to the RTL design. We added the bit-level support into the symbolic execution engine. First, we need to inform the engine the number of bits for a variable as the third parameter in the make_symbolic function. As shown in Figure 5, we specify 1 as the third parameter as the size of variable *sel*. In this way, we can generate correct test cases: *sel* = 0 and *sel* = 1.

### IV. PRELIMINARY RESULTS

We have developed a prototype of our generic symbolic execution framework for RTL design validation. Our test generation tool takes a Verilog design and a test harness as inputs and produces test cases. We have modified KLEE symbolic execution engine to support bit-level variables and handle some C++ features. In this section, we present the preliminary results of our case studies. All experiments were performed on 2.5GHz Intel i5 Processor with 4GB memory.

### A. The Overview of FPU Design

The FPU design is a double precision floating point core [32]. The core is designed to meet the IEEE 754 standard [33] for double precision floating point arithmetic. The core supports four operations: add, subtract, multiply and divide. The FPU design is an available project in the OpenCore website [34].

The FPU design takes two 64 bits input operands, the operation code and the rounding mode as the inputs. After the computation, it produces one 64 bits' output.

The basic hierarchy of the FPU design is shown in Figure 6. There are six modules: fpu_add, fpu_sub, fpu_mul, fpu_div, fpu_round and fpu_exceptions. The design takes the inputs and conduct the operations. After the operation, the rounding is conducted and the exception is checked. Then the output is produced.

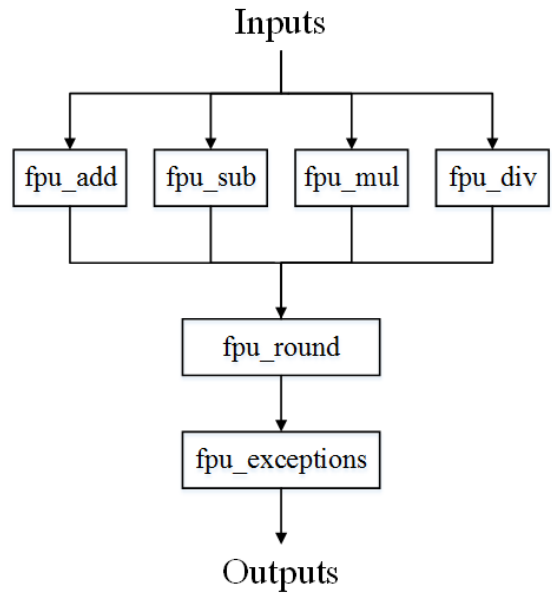

Fig. 6 The hierarchy of FPU design.

*B. Generation Results*

To symbolically execute the FPU design, we created a test harness and compile the harness and the design into a C++ program. Then we employed the symbolic execution engine to run the design.

The symbolic execution engine can be affected by some system factors like memory usage and CPU usage. We run symbolic execution on FPU designs many times to check if we can get the similar results every time. The experimental results prove that the number of generated tests are almost the same during different iterations of symbolic execution. We summarized the number of generated tests, the number of generated test vectors and time cost in Table 1.

TABLE 1. THE NUMBER OF GENERATED TESTS

| Tests(#) | Test Vectors(#) | Time(Min) |
|---|---|---|
| ~4010 | ~12060 | ~17 |

We also collected the generation overhead data in Table 2. The result shows that symbolic execution of FPU design consumed reasonable memory and CPU.

TABLE 2. THE CPU AND MEMORY USAGE

| Memory(%) | Memory(Mb) | CPU(%) |
|---|---|---|
| 15.3% | 627 | 40 |

With generated test cases, we further created a test bench and applied the generated test cases using RTL simulation. With RTL simulation, we can better evaluate the coverage of generated tests and observe the behavior of the FPU design upon the generated tests. Here we employed Mentor Graphics Modelsim [35] as our RTL simulation tool to run the simulation.

To better observer the design behavior, the developers can check all singals triggerred by the generated test cases. The simulation wave is shown in Figure 7.

We further evaluated the coverage of generated tests on FPU design, the coverage result is shown in Figure 8. From the result, we can find that the generated tests can achieve 98.4% and 96.8% on statement and branch separately. FPU design includes a lot of complex logics. Our SE4RDV framework generated high quality tests and triggered most functionalities on FPU design.

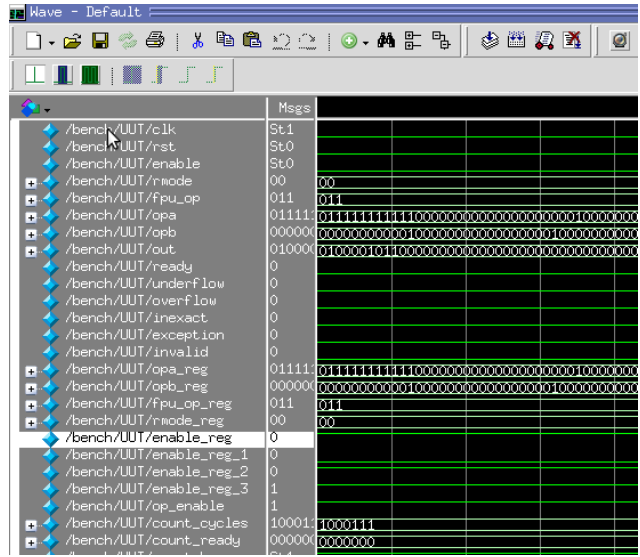

Fig. 7 The simulation wave output.

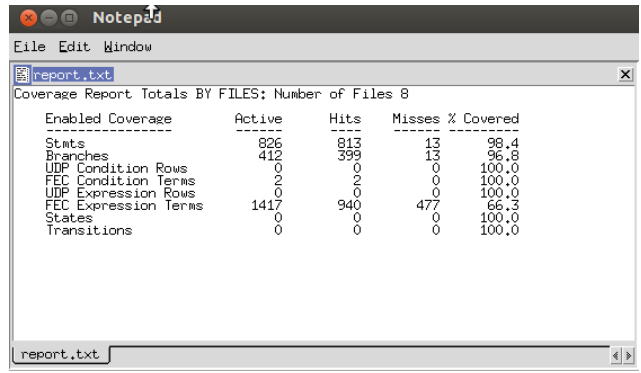

Fig. 8 The overall coverage result.

V. CONCLUSIONS

In this paper, we present a generic symbolic execution framework for RTL designs by reusing available software tools. This approach only requires minimum effort to develop a test harness for each RTL design. After applying our approach to a FPU design, the experimental results show that our approach generates high-quality test cases to achieve high coverage. In the future, we will apply this approach to more RTL designs. Moreover, we will try to propose a new generation strategy to eliminate unnecessary generated tests based on the hardware nature.